# Inverse Design of Integrated Terahertz Vortex Beam Emitters with Staged-Annealing Topology Optimization

Faqian Chong[1, 2, 3, †], Tiancheng Zhang[1, 2, 3, †], Yulun Wu[1, 2, 3], Bingtao Gao[1, 2, 3], Yingjie Wu[1, 2, 3], Shilong Li[1, 2, 3], Hongsheng Chen[1, 2, 3, 4, *], Song Han[1, 2, 3, *]

[1]Innovative Institute of Electromagnetic Information and Electronic Integration, College of Information Science & Electronic Engineering, Zhejiang University, Hangzhou, 310027, China

[2]State Key Laboratory of Extreme Photonics and Instrumentation, ZJU-Hangzhou Global Scientific and Technological Innovation Center, Zhejiang University, Hangzhou 310200, China

[3]International Joint Innovation Center, The Electromagnetics Academy at Zhejiang University, Zhejiang University, Haining 314400, China

[4]Key Lab. of Advanced Micro/Nano Electronic Devices & Smart Systems of Zhejiang, Jinhua Institute of Zhejiang University, Zhejiang University, Jinhua 321099, China

*Correspondence to: hansomchen@zju.edu.cn, song.han@zju.edu.cn.

## Abstract

Integrated photonics is increasingly demanded in applications such as large-scale data centers, intelligent sensing, and next-generation wireless communications, where compact, multifunctional, and energy-efficient components are essential. Inverse-designed photonics, empowered by optimization and learning algorithms, have emerged as a powerful paradigm for realizing compact and multifunctional integrated photonic components. In this work, we develop a staged-annealing topological optimization (SATO) framework tailored for the design of integrated terahertz (THz) beam-shaping devices. Employing this inverse-designed framework, we experimentally demonstrate a class of compact THz vortex beam emitters on an all-silicon on-chip platform. These devices efficiently convert the in-plane fundamental transverse electric (TE) waveguide mode into free-space vortex beams with mode purity up to 87% and energy conversion efficiency up to 74% across the target wavelength range (680 μm to 720 μm). The inverse-designed emitters exhibit ultracompact footprints (lateral size < 4λ) and a free-standing configuration, enabling the generation of dual-directional vortex beams carrying opposite topological charges. The proposed SATO framework provides a generalizable and fabrication-compatible approach for THz photonic device engineering, offering a scalable pathway toward complex structured beam manipulation in next-generation wireless communication systems and on-chip integrated THz photonic systems.

# 1 introduction

Inverse design of photonic devices based on optimization and learning algorithms has recently emerged as a highly promising paradigm in photonics. Compared with traditional forward design approaches that rely on physical intuition and parameter sweeping, inverse design enables automatic optimization in high-dimensional structural space by constructing objective functions tailored to specific optical functionalities. This greatly expands both the design freedom and performance boundaries of photonic devices[1,2]. To this end, researchers have proposed various efficient algorithms, including direct binary search (DBS)[3-5], topology optimization (TO)[6-9], particle swarm optimization (PSO)[10,11], and neural networks (NN)[12-15], which have been widely applied to the high-performance design of diverse photonic components such as mode converters[16], mode demultiplexers[17,18], optical logic gates[19,20], and grating couplers[21,22]. However, these optimization strategies are typically single-stage and often suffer from premature convergence or entrapment in local optima, particularly when dealing with the complex inverse design tasks involving structured light fields, which frequently require satisfying multiple objectives under strong constraints.

To address this challenge of optimizing complex structured light fields, an advanced and efficient inverse-design algorithm is required. In this work, we propose a staged-annealing topology optimization (SATO) method, and demonstrate its capabilities by applying it to the inverse design of orbital angular momentum (OAM) beam emitters operating in the terahertz (THz) region. The OAM beam, or vortex beam, characterized by its helical phase structure and mode orthogonality, has demonstrated wide-ranging value in optics and electromagnetic applications. The optical OAM modes carrying different topological charges (OAM orders) support multi-channel parallel communication, significantly enhancing information capacity[23,24], while also finding broad applications in super-resolution imaging[25,26], optical manipulation, quantum information, and metrology[27,28]. Particularly in the THz regime, the combination of high-dimensional orthogonality and abundant spectral resources of OAM modes provides crucial support for building high-capacity and high-speed free-space communication systems[29,30]. Nevertheless, realizing the efficient, compact, and manufacturable OAM devices in the THz band remains challenging. Conventional methods for OAM generation, including spiral phase plates[31,32], diffraction gratings[33,34] metasurfaces[35-40], and digital subwavelength structures[41], have achieved breakthrough advances but often suffer from limitations in footprint, operational bandwidth, or fabrication complexity, making it difficult to meet the combined requirements of broadband, efficient radiation and integration for THz photonic systems.

Leveraging the proposed SATO method, we designed and fabricated a family of highly integrated THz photonic OAM emitters on high-resistivity silicon wafer. Experimental validation was performed using a vector-network-analyzer-based THz near-field scanning system, which measured both magnitude and phase profiles of the emitted fields. The results demonstrate robust and broadband emission of vortex beams with topological charges ranging from ±1 to ±3, achieving total radiation efficiencies up to 74% and mode purities exceeding 87% across the designated frequency band. These findings validate the effectiveness of the SATO framework in delivering high-performance THz photonic devices tailored for structured light generation. Furthermore, the proposed optimization paradigm is broadly applicable and can be extended to the inverse design of other complex beam-shaping devices requiring simultaneous control of multiple optical degrees of freedom.

# 2 Results

## 2.1 Design Principle of Terahertz Photonic Vortex Emitter

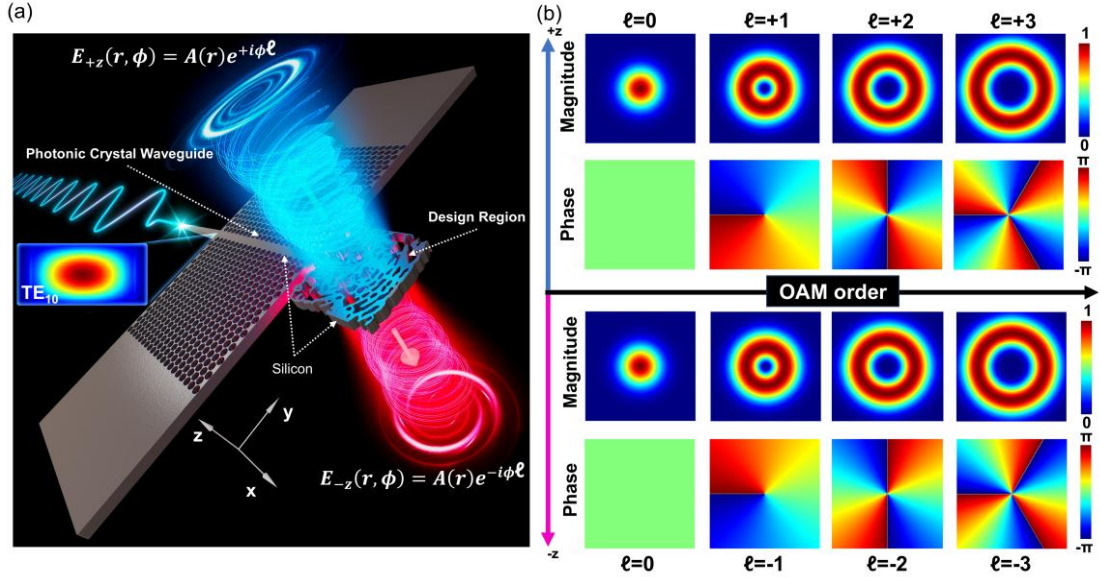

**Figure 1. Concept of an Inverse-Designed THz Photonic Vortex Beam Emitter.** (a) Schematic of the inverse-designed THz photonic vortex beam emitter. A fundamental TE mode ($TE_{10}$) is coupled into a photonic crystal (PhC) waveguide via a tapered coupler, which subsequently excites the inverse-designed region to radiate vortex beams of opposite topological charges ($\ell$) along the ±z-directions. (b) Cross-sectional views of vortex beams carrying OAM orders (i.e., topological charges) from -3 to 3, showing both the magnitude and phase distributions.

The schematic of the proposed THz photonic vortex emitter is shown in Figure 1a, where the functional design region is patterned on a suspended high-resistivity silicon wafer. To enable efficient coupling of the in-plane THz wave into the vortex-emitting structure (i.e., the inverse-designed region), a tapered waveguide is employed. Positioned between the taper and vortex emitter, a line-defect waveguide in photonic crystal (PhC) slab is strategically incorporated to fulfill multiple critical functions: (1) it provides mechanical support for both the tapered input waveguide and the inverse-designed vortex emitter; (2) it exhibits a photonic bandgap within the target frequency range, which suppresses lateral mode leakage through interference effects, thereby confining the THz waves within the central guiding region[42]; and (3) to facilitate device fabrication and experimental handling, a bare silicon region is introduced laterally beyond the PhC slab, which also benefits potential large-scale device integration. It is worth noting that, owing to its free-standing configuration, the vortex emitter exhibits mirror symmetry along the z-axis. As a result, it supports dual-channel emission, wherein the emitted OAM beams propagate in opposite directions with the same order but opposite topological charges, as shown in Figure 1a. The emitted beams from the device exhibit the field profiles of Laguerre-Gaussian (LG) vortex modes, which are expressed as[43]:

$$E_{p,\ell}(r,\phi) = \frac{1}{w_0} L_p^{|\ell|}\left(\frac{2r^2}{w_0^2}\right) \sqrt{\frac{2p!}{\pi(p+|\ell|)!}} \left(\frac{\sqrt{2}r}{w_0}\right)^{|\ell|} e^{-r^2/w_0^2} e^{i\ell\phi} \qquad (1)$$

where $(r, \phi)$ are the spatial coordinates and phase, $w_0$ denotes beam waist, $p$ is the radial index, and $\ell$ is the azimuthal (topological) charge. The phase term $\phi$ accumulates linearly around the azimuthal direction and integrates to $2\pi\ell$. The parameters $w_0$ and $p$ are the radial index, respectively. In this work, we focus on the case of first-order radial index ($p = 1$), a commonly adopted simplification, leading to the reduced LG-vortex expression:

$$E_\ell(r, \phi) = \frac{1}{w_0} \sqrt{\frac{1}{\pi |\ell|!}} \left(\frac{\sqrt{2}r}{w_0}\right)^{|\ell|} e^{-r^2/w_0^2} e^{i\ell\phi} \qquad (2)$$

Representative examples of these LG beams with topological charges $\ell = 0, 1, 2, 3$ are illustrated in Figure 1b, corresponding to upward emission along the +z-direction. The opposite signs $(-\ell)$ are emitted simultaneously in the -z-direction due to the mirror symmetry of the suspended device.

## 2.2 Inverse Design Framework of SATO

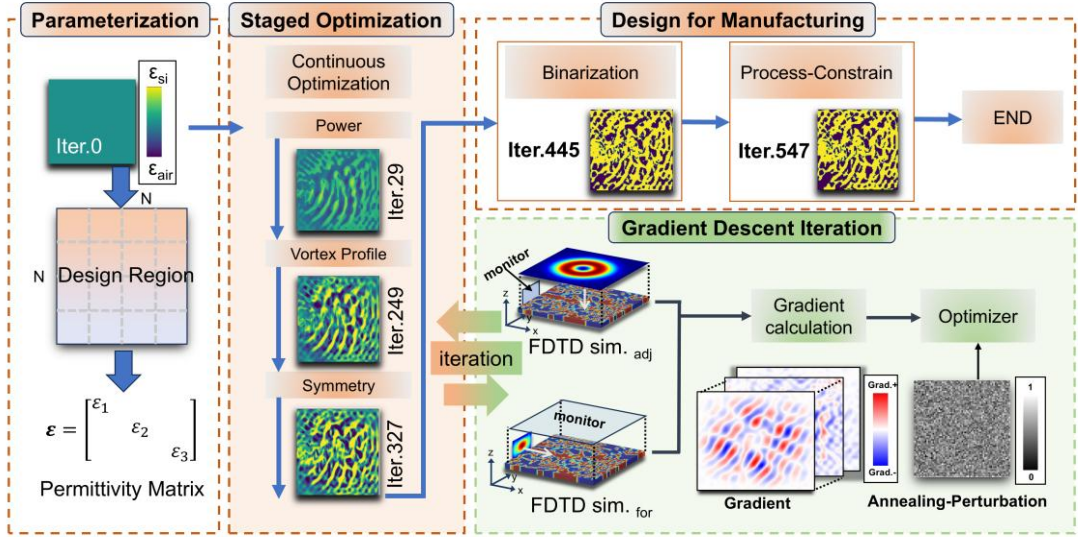

**Figure 2. Flowchart of the SATO inverse design framework.** The SATO method comprises three core sections: parameterization, staged optimization, and design for manufacturing. The staged optimization section employs an adjoint-based annealing gradient descent algorithm.

In conventional schemes, applying complex and stringent objective functions at the initial stage frequently leads to premature convergence due to sparse or ill-conditioned gradient information, thereby constraining the improvement of overall device performance[44]. To mitigate this, we decompose the optimization process into hierarchical stages, forming a "coarse-to-fine" progression procedure. Additionally, controlled perturbations are introduced throughout the continuous optimization to enhance the algorithm's ability to escape local minima and expand the search space.

Figure 2 illustrates the flowchart of the SATO framework, consisting of three main sections: parameterization, staged optimization, and design for manufacturing (see Supplementary Note S1 for details). In the parameterization, the design region is discretized into N×N pixels, and a two-dimensional (2D) permittivity matrix $\varepsilon$ is constructed to describe material properties for each pixel, serving as the foundation for parameter optimization. The initial design uses a binary silicon-air material configuration, with an average permittivity assigned to all pixels as the starting point. The device is designed to achieve low insertion loss and vortex beam emission with high mode purity within an ultra-compact footprint. To achieve this, we define the optimization objective as maximizing the figure of merit (FOM) associated with the design variables. The inverse design problem is mathematically formulated as:

$$\max_{\varepsilon} FOM(\varepsilon) \tag{3}$$

$$s.t. \nabla \times \frac{1}{\mu} \nabla \times E(\varepsilon) - \omega^2 \varepsilon E(\varepsilon) = -i\omega J \tag{4}$$

where $\mu$ is the magnetic permeability, $\varepsilon$ is the spatial permittivity distribution, $E$ is the electric field, $\omega$ is the angular frequency, and $J$ represents the excitation current source[45]. The second section is staged optimization, which allows the permittivity within the design region to vary continuously between those of silicon and air. Specifically, the initial stage of optimization sets a relatively simple performance objective to guide the formation of a basic functional structure. Subsequently, more complex physical constraints and performance metrics are gradually introduced, enabling a stepwise evolution of the FOM. This strategy enhances the global search capability and reduces the risk of being trapped in local minima, thus ensuring high-performance device realization[46]. In practice, different FOMs are applied at different stages, arranged to increase optimization complexity. In the first stage, the FOM focuses solely on maximizing emission efficiency. The FOM is defined as follows:

$$FOM_1 = \frac{\int_{S_{out}} Re(E \times H^*) \cdot dS}{\int_{S_{in}} Re(E \times H^*) \cdot dS} \tag{5}$$

where the numerator represents the power flux on the dual-output plane of the vortex beam, and the denominator represents the power flux on the input plane of the waveguide. This formula maximizes the emission efficiency along the positive and negative halves of the z-axis. Building on this, the second stage introduces the phase structure and modal characteristics of the target vortex beam by modifying the FOM accordingly:

$$FOM_2 = \gamma \frac{|\int (E \times H_m) \cdot dS + \int (E_m \times H) \cdot dS|^2}{\int Re(E_m \times H_m) \cdot dS} \tag{6}$$

$E_m$ and $H_m$ represent the field distributions of the target vortex beam, while $E$ and $H$ are the actual fields obtained from direct simulation, and $\gamma$ is a normalization constant. Finally, to ensure the symmetric mode profile and wavefront quality of the emitted vortex beams, the symmetry constraints are incorporated into the second-stage FOM to form the third stage refined optimization objective:

$$FOM_3 = FOM_2 - \gamma_{sym} Loss_{sym} \tag{7}$$

$$Loss_{sym} = \int_{S_{out}} \left||E(r)| - |E_{sym}(r)|\right|^2 \cdot dS \tag{8}$$

where $|E(r)|$ is the electric field magnitude distribution within one quadrant sector of the output plane, and $|E_{sym}(r)|$ denotes the magnitude distribution of the other three symmetric sectors. The optimization goal is to make the magnitude distributions of the four sectors as consistent as possible to achieve the four-fold symmetry of the annular beam.

The iterative optimization process described above adopts an annealing-based gradient descent method, which is built upon the adjoint method (see Supplementary Note S2 for more details)[47]. During the continuous optimization, Gaussian perturbations are introduced in each iteration to the permittivity matrix $\varepsilon$, employing a simulated annealing-like perturbation mechanism[48]. This enhances the ability of the optimization to escape local minima and improves the final performance of the structure. To ensure compatibility with actual manufacturing processes, a design for manufacturing section is introduced after continuous optimization. In this section, the continuous permittivity distribution is binarized by using a Heaviside filter, and a β-ramping strategy is applied to gradually transition the structure from smooth grayscale to a silicon-air binary configuration. However, simple binarization may lead to fabrication-rule violations, such as fine lines or sharp corners smaller than the minimum feature size. To address this issue, a process-constrained optimization section is further introduced[49]. By incorporating penalty terms, the structure is guided to avoid unmanufacturable features while preserving performance, thereby ensuring a final design that balances both optical performance and fabrication feasibility.

## 2.3 Device Fabrication and Experimental Methods

The inverse-designed THz photonic devices were fabricated using standard photomask lithography to define the photonic pattern on a thick photoresist layer (AZ4620, ~7 μm). The pattern was directly transferred onto the silicon substrate using deep reactive ion etching (DRIE)[50]. The fabrication details can be found in Supplementary Note S3. The devices were patterned on a high-resistivity (>10,000 Ω·cm) single-crystal silicon wafer with a thickness of 200 μm. Figure 3a presents the optical microscope image of a representative fabricated device, which includes a tapered waveguide coupler, a line-defect waveguide with a width of 210 μm embedded in a photonic crystal slab exhibiting a TE-like bandgap (see Supplementary Note S4 for details)[42], and the inverse-designed vortex beam emitter. In addition to the device layout, the calculated photonic band diagram of the photonic crystal unit cell is also shown, highlighting the bandgap characteristics and mode confinement relevant to guided mode propagation. The photonic crystal waveguide is engineered for robust guided mode propagation, whose dispersion relation is illustrated alongside the corresponding photonic band structure. To experimentally verify the performance of the proposed SATO algorithm, four vortex beam emitters with target topological charges $\ell = 0, 1, 2, 3$ were designed, fabricated, and characterized. For the devices targeting topological charges from $\ell = 0$ to $\ell = 2$, the inverse-designed region has a lateral footprint of 2.4 mm × 2.4 mm. The device designed for the $\ell = 3$ vortex beam requires an expanded inverse-designed area of 4.2 mm × 4.2 mm. This increase in size is necessitated by the intrinsic characteristics of the LG vortex beams, where higher-order modes exhibit broader intensity rings, requiring a larger aperture to maintain beam fidelity and high mode purity. All inverse-designed vortex emitters exhibit a freeform geometry (right column in Figure 3a), with the smallest feature size exceeding 10 μm. This ensures compatibility with DRIE through the 200 μm-thick silicon slab, and maintains a fabrication aspect ratio larger than 10:1. To enforce this constraint during the design stage, a filter function is integrated into the SATO algorithm, allowing precise control over the minimum feature size and etch-friendly geometry, thereby ensuring both fabrication feasibility and structural integrity of the final device.

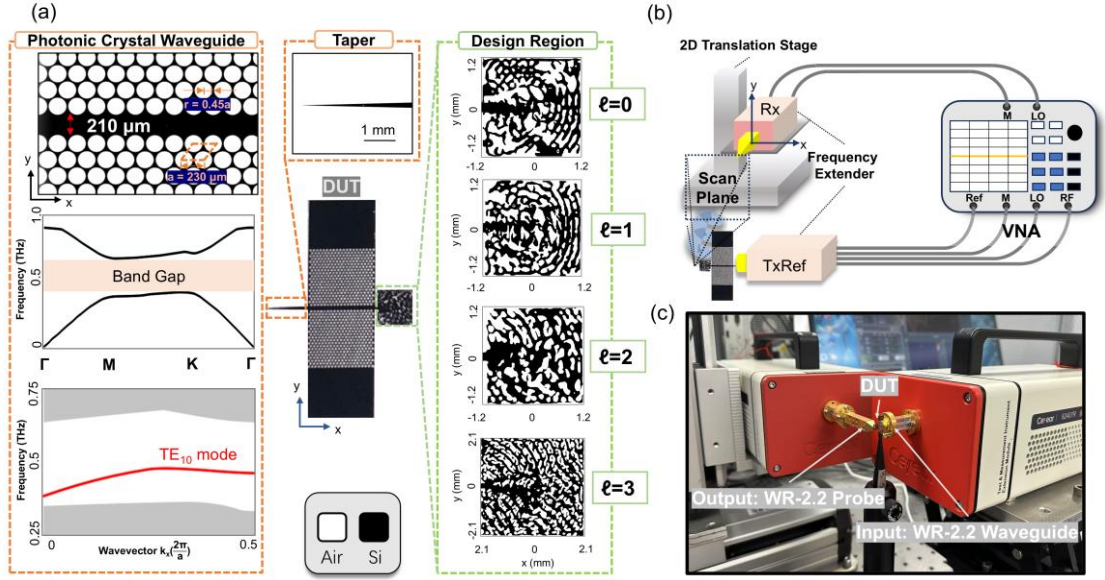

**Figure 3. Fabricated device and experimental characterization setup.** (a) Microscope image of the fabricated inverse-designed vortex beam emitter (device under test, or DUT), integrated with a PhC line-defect waveguide and a topologically structured design region. The calculated photonic band diagram of the unit cell is presented to highlight the mode confinement and bandgap characteristics. The line defect in the PhC slab features a waveguide engineered for robust guided mode propagation, whose dispersion relation is illustrated alongside the corresponding photonic band structure. Distinct pattern geometries within the design region encode different OAM states. (b) Schematic of the experimental setup, illustrating the THz continuous-wave (CW) measurement system composed of a vector network analyzer (VNA) and frequency extension modules covering 330-500 GHz. A scanning field probe in the XY plane was employed to record the magnitude and phase distributions of the emitted electric field, with signals transmitted through a separate extension module. (c) Photograph of the field measurement setup. The extension module is connected to the device via a WR-2.2 waveguide, while the emitted signal is collected by a WR-2.2 waveguide probe and sent to the receiver module.

To experimentally characterize the performance of the inverse-designed vortex beam emitters, a custom-built 2D field scanning system was developed, integrated with vector network analyzer extension (VNAX) modules. A schematic of the measurement setup is presented in Figure 3b, with a photograph of the actual system shown in Figure 3c. The THz generation-detection setup consists of two electronic frequency extension modules-one configured as a Transmit-Reference (TxRef) module and the other as a Receive-only (Rx) module-interfaced with a four-port vector network analyzer (VNA), which simultaneously serves as the local oscillator (LO), radio frequency (RF) source, and spectrum analyzer (see Supplementary Note S5 for further details).

During measurements, the TxRef and Rx modules were connected to the device under test (DUT) via a WR-2.2 waveguide and an orthogonally arranged tapered WR-2.2 probe to enable efficient excitation and field detection. The DUT was fixed with a vertical sample holder with its tapered coupler inserted into the WR-2.2 TxRef waveguide. The Receive-only (Rx) probe was mounted on a high-precision motorized 2D translation stage, allowing controlled scanning in the XY plane to simultaneously visualize the spatial field intensity and phase of the emitted vortex beams. It is worth emphasizing that the experimental setup enables direct characterization of both the spatial magnitude and the phase profile of the emitted beams. This eliminates the need for conventional self-interference techniques typically employed in optical systems for OAM detection[43], thus providing a more straightforward and compact platform for THz vortex beam analysis.

## 2.4 Device Performance

We conducted full-wave simulations for the inverse-designed vortex beam emitters using Ansys Lumerical FDTD solver. Figure 4a presents the simulated magnitude (up panel) and phase (down panel) distributions for each order of the devices. The vortex beam carrying the same topological charge order with opposite sign is taken from one inverse-designed device by setting the monitor plane above (+z) and below (-z) the emitter. Figure 4b shows the experimentally measured field magnitude and phase distributions. The experimental results agree well with the numerical simulations, and both clearly demonstrate that the inverse-designed THz photonic devices can efficiently emit high-quality vortex beams. Although the measured OAM beams agree very well with the numerical simulations, the experimental results exhibit some distinguishable discrepancies, resulting in a less distinct minimum magnitude in the vortex beam center and some fluctuations in overall intensity. This phenomenon may be attributed to three main factors: First, the numerical simulation involves idealized situation, without considering insertion loss, material loss, material surface roughness caused scattering loss, and so on. However, the actual fabrications and experiments introduce inevitable losses that may prevent the excitation of a pure quasi-TE mode in the tapered waveguide coupler, thus affecting the beam quality. Second, despite the incorporation of fabrication-aware constraints during the design process, certain intricate geometrical features may still pose fabrication challenges, leading to structural imperfections or defects during manufacturing. Finally, during measurement, minor positioning errors in the 2D scanning stage may cause deviations between the measured electric field intensities and the simulation results.

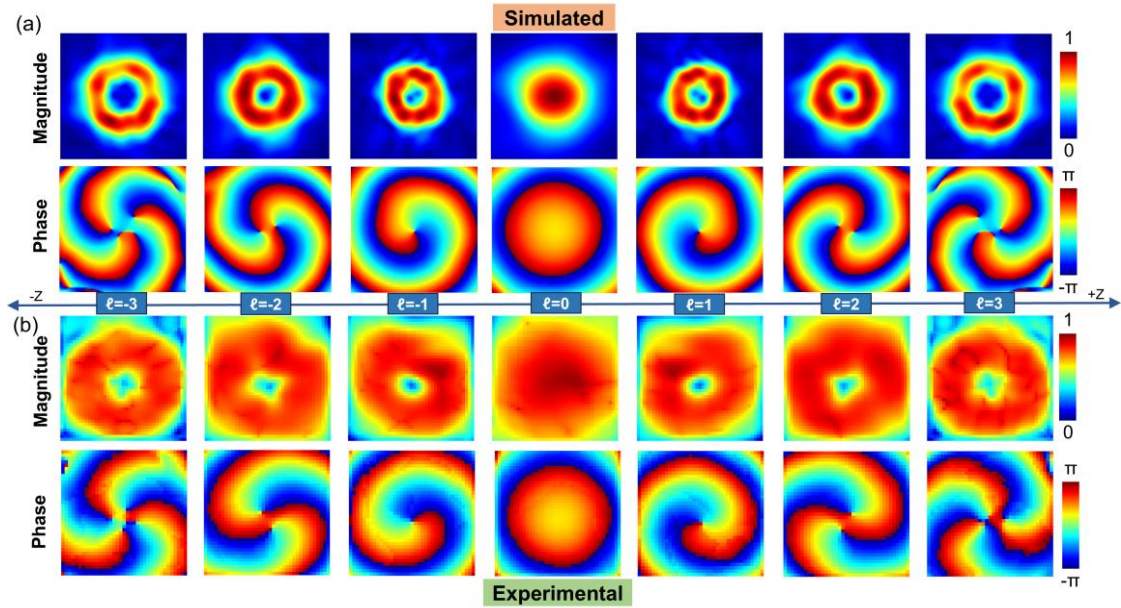

**Figure 4. Magnitude and phase distributions of vortex beams at center wavelength of 700 μm for different topological charges.** (a) Simulated results. (b) Experimental results.

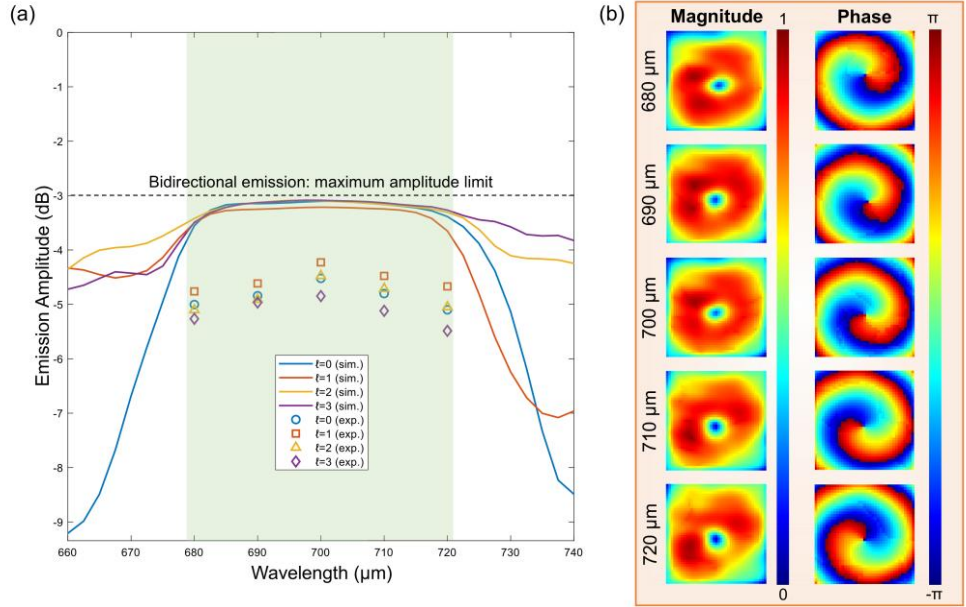

**Figure 5. Broadband performance of the device.** (a) Comparison of the emission efficiency between simulated and experimental results, highlighting consistent broadband performance. (b) Measured magnitude and phase distributions of the +1-vortex beam across the operational bandwidth, demonstrating the broadband characteristics of the inverse-designed THz photonic devices.

Figure 5a displays the simulated and experimental efficiencies of single-direction vortex beam radiation for the four orders of devices within the operating bandwidth (see Supplementary Notes S6 and S7 for further details). Due to the bidirectional emission characteristics of the devices, the theoretical maximum efficiency for a single direction is 50% (-3 dB), which serves as the reference for all efficiency calculations presented in this study. From the results, it can be observed that the simulated efficiencies of the devices remain within the range of -3 to -4 dB across the bandwidth, while the best experimental efficiency reaches up to -4.2 dB for the +1st-order device at 700 μm. The maximum deviation between simulated and experimental efficiencies occurs for the 3rd-order device at an operating wavelength of 720 μm, with a difference of approximately 2.2 dB. The experimental results indicate that all four orders of the designed devices can efficiently radiate vortex beams within a certain operating bandwidth, demonstrating robust performance and stable field emission.

Additionally, Figure 5b further provides the measured magnitude and phase distributions of the +1st-order vortex beam emitter within its operating bandwidth. It is evident that the vortex beam is consistently and clearly generated from 680 μm to 720 μm, as reflected by the well-formed doughnut-shaped intensity profiles and the spiral-like phase fronts. As a representative example, this clearly illustrates the device's ability to maintain high effciency within the target band, which is essential for practical applications such as broadband communication and THz imaging.

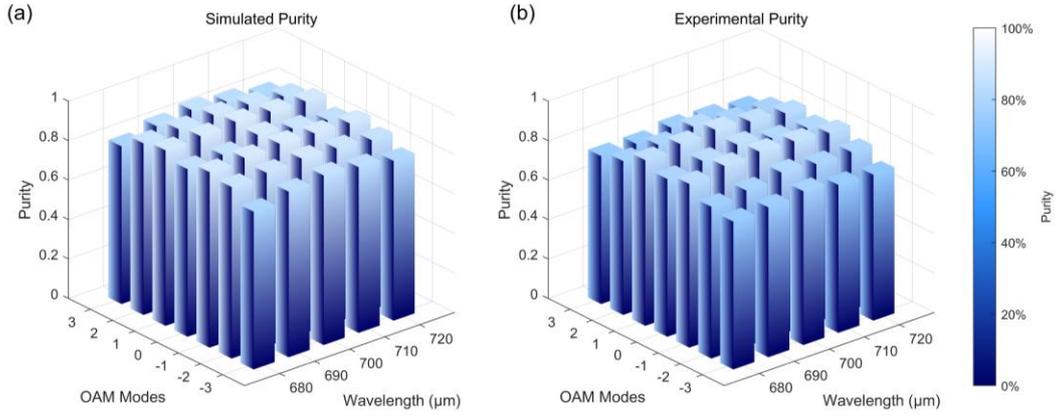

**Figure 6. OAM mode purity of the broadband vortex beam emitters.** (a) Simulated OAM mode purity of devices with different topological charges over a range of frequencies. (b)Measured OAM mode purity at corresponding frequencies, demonstrating broadband and high-purity performance across multiple OAM states.

The purity of the OAM mode is a key parameter for evaluating the performance of a vortex beam emitter, characterizing the proportion of a specific OAM mode within the overall radiation field. We employ a Fourier transform method to analyze the angular field distribution of the emitted beam, thereby extracting the intensity of each OAM mode component and calculating the purity of the target mode[51]. The electric field distribution on the observation plane at a sampling radius $r$ is denoted as $E(\phi)$, where $\phi$ is the polar coordinate angle. The OAM mode component $A_\ell$ can be obtained using the following formula:

$$A_\ell = \frac{1}{2\pi} \int_{-\pi}^{\pi} E(\phi) \cdot e^{-i\ell\phi} \, d\phi \qquad (9)$$

This formula represents the Fourier transform expanded in angular coordinates, where $\ell$ denotes the angular momentum order. The inverse transform can be written as:

$$E(\phi) = \sum_{\ell} A_\ell e^{i\ell\phi} \qquad (10)$$

where $A_\ell$ is the complex amplitude coefficient corresponding to each OAM mode $\ell$. Further, for a specific OAM mode of order $\ell_0$, the mode purity is defined as the ratio of the energy of that mode to the total energy of all modes:

$$Purity(l_0) = \frac{|A_{\ell_0}|^2}{\Sigma_{\ell=\ell_a}^{\ell_b}|A_\ell|^2} \qquad (11)$$

where $[\ell_a, \ell_b]$ represents the range of OAM mode orders considered in the analysis. In this work, the range is typically taken as $\ell \in [-5,5]$. This method is applicable to both simulation and experimental measurement data and can be used to evaluate the stability and orthogonality of OAM modes under different frequencies or polarization conditions.

Figures 6a and 6b present the extracted mode purities for each topological order at various wavelengths. The experimentally measured OAM purities show good agreement with the numerical simulations, confirming that the generated vortex beams maintain high mode purity across the operational bandwidth. For each device, the mode purity peaks near its designated central wavelength, indicating effective suppression of unwanted OAM components. Across all measured devices and wavelengths, the mode purity consistently exceeds 75%, with the highest recorded value reaching approximately 87%. It should also be noted that, while the experimentally demonstrated maximum topological charge is $\ell = 3$, the proposed inverse-design algorithm is, in principle, capable of generating vortex beams with arbitrary OAM orders.

## 3 Conclusion

This work introduces a generalizable and fabrication-compatible inverse design framework, i.e., the staged-annealing topological optimization (SATO), which enables the precise and efficient realization of complex structured light fields. By integrating staged-optimization with process-constraints, the SATO framework provides a robust and scalable strategy for designing high-performance photonic structures under multi-objective and fabrication-constrained conditions. To validate its effectiveness, the SATO approach is applied to design compact, all-silicon integrated THz photonic vortex beam emitters, carrying arbitrarily defined topological charges with high OAM mode purity and high conversion efficiency (A comparison of vortex beam emitters with other designs can be found in Supplementary Note S8). The inverse-designed devices are co-designed with photonic crystal waveguides, which facilitate efficient in-plane THz excitation, mechanical support, and seamless compatibility with future modular integration and array-level packaging. Experimental characterization of the fabricated devices exhibits well-defined field magnitudes and phase profiles over a broad frequency range, with experimental measurements agreeing well with the numerical simulations. These findings position the SATO framework as a powerful tool for advancing scalable THz on-chip photonics with potential impact across communications, imaging, and sensing.

**Supplementary Information**
The online version contains supplementary material available online

**Acknowledgments**
S. H. acknowledges the National Key R&D Program of China (2024YFB2808200), the National Natural Science Foundation of China (62475230), the Excellent Young Scientists Fund Program (Overseas) of China, and the Fundamental Research Funds for the Central Universities.

**Authors' Contributions**
S. H initiated the idea. F. C developed the algorithm and performed the simulations. T. Z and Y. L. W assisted in the device design and conducted the experiments. B. G and Y. J. W supervised and guided the experimental procedures. F. C and S. H wrote the manuscript. H. C and S. H supervised the entire project. All authors discussed the results and contributed to the final manuscript.

**Data availability**
The data that support the findings of this work are available from the corresponding author upon reasonable request.

**Competing interest**
The authors declare no competing interests.

# References

1. Lu, J. & Vučković, J. Nanophotonic computational design. *Opt. Express* **21**, 13351-13367, doi:10.1364/OE.21.013351 (2013).

2. Molesky, S. *et al.* Inverse design in nanophotonics. *Nature Photonics* **12**, 659-670, doi:10.1038/s41566-018-0246-9 (2018).

3. Wang, Q., Chumak, A. V. & Pirro, P. Inverse-design magnonic devices. *Nature Communications* **12**, 2636, doi:10.1038/s41467-021-22897-4 (2021).

4. Li, Z. *et al.* Ultra-compact high efficiency and low crosstalk optical interconnection structures based on inverse designed nanophotonic elements. *Scientific Reports* **10**, 11993, doi:10.1038/s41598-020-68936-w (2020).

5. Xu, Y. *et al.* Inverse-designed ultra-compact high efficiency and low crosstalk optical interconnect based on waveguide crossing and wavelength demultiplexer. *Scientific Reports* **11**, 12842, doi:10.1038/s41598-021-92038-w (2021).

6. Udupa, A., Zhu, J. & Goddard, L. L. Voxelized topology optimization for fabrication-compatible inverse design of 3D photonic devices. *Opt. Express* **27**, 21988-21998, doi:10.1364/OE.27.021988 (2019).

7. Chen, Y. *et al.* Topology Optimization-Based Inverse Design of Plasmonic Nanodimer with Maximum Near-Field Enhancement. *Advanced Functional Materials* **30**, 2000642, doi:https://doi.org/10.1002/adfm.202000642 (2020).

8. Colburn, S. & Majumdar, A. Inverse design and flexible parameterization of meta-optics using algorithmic differentiation. *Communications Physics* **4**, 65, doi:10.1038/s42005-021-00568-6 (2021).

9. Min Chen, Lian Shen, Yifei Hua, Zijian Qin & Wang, H. Topology-optimized Plasmonic Nanoantenna for Efficient Single-photon Extraction. *Progress In Electromagnetics Research* **180**, 55-60 (2024).

10. He, W. *et al.* Ultrafast all-optical terahertz modulation based on an inverse-designed metasurface. *Photon. Res.* **9**, 1099-1108, doi:10.1364/PRJ.423119 (2021).

11. Kokhanovskiy, A. *et al.* Inverse design of mode-locked fiber laser by particle swarm optimization algorithm. *Scientific Reports* **11**, 13555, doi:10.1038/s41598-021-92996-1 (2021).

12. Ma, W. *et al.* Deep learning for the design of photonic structures. *Nature Photonics* **15**, 77-90, doi:10.1038/s41566-020-0685-y (2021).

13. Yuan, M. *et al.* Inverse design of a nano-photonic wavelength demultiplexer with a deep neural network approach. *Opt. Express* **30**, 26201-26211, doi:10.1364/OE.462038 (2022).

14. Kim, J. *et al.* Inverse design of an on-chip optical response predictor enabled by a deep neural network. *Opt. Express* **31**, 2049-2060, doi:10.1364/OE.480644 (2023).

15. Tanriover, I., Lee, D., Chen, W. & Aydin, K. Deep Generative Modeling and Inverse Design of Manufacturable Free-Form Dielectric Metasurfaces. *ACS Photonics* **10**, 875-883, doi:10.1021/acsphotonics.2c01006 (2023).

16. Callewaert, F., Butun, S., Li, Z. & Aydin, K. Inverse design of an ultra-compact broadband optical diode based on asymmetric spatial mode conversion. *Scientific Reports* **6**, 32577, doi:10.1038/srep32577 (2016).

17. Chang, W. *et al.* Ultra-compact mode (de) multiplexer based on subwavelength asymmetric Y-junction. *Opt. Express* **26**, 8162-8170, doi:10.1364/OE.26.008162 (2018).

18  Di Domenico, G., Weisman, D., Panichella, A., Roitman, D. & Arie, A. Large-Scale Inverse Design of a Planar On-Chip Mode Sorter. *ACS Photonics* **9**, 378-382, doi:10.1021/acsphotonics.1c01539 (2022).

19  Qingze Tan, Chao Qian & Chen, H. Inverse-designed Metamaterials for On-chip Combinational Optical Logic Circuit. *Progress In Electromagnetics Research* **176**, 55-65 (2023).

20  Neşeli, B., Yilmaz, Y. A., Kurt, H. & Turduev, M. Inverse design of ultra-compact photonic gates for all-optical logic operations. *Journal of Physics D: Applied Physics* **55**, 215107, doi:10.1088/1361-6463/ac5660 (2022).

21  Piggott, A. Y. *et al.* Inverse design and implementation of a wavelength demultiplexing grating coupler. *Scientific Reports* **4**, 7210, doi:10.1038/srep07210 (2014).

22  Su, L. *et al.* Fully-automated optimization of grating couplers. *Opt. Express* **26**, 4023-4034, doi:10.1364/OE.26.004023 (2018).

23  Yan, Y. *et al.* High-capacity millimetre-wave communications with orbital angular momentum multiplexing. *Nature Communications* **5**, 4876, doi:10.1038/ncomms5876 (2014).

24  Li, X. *et al.* Resolution Analysis of Coincidence Imaging Based on OAM Beams With Equal Divergence Angle. *IEEE Transactions on Antennas and Propagation* **71**, 2891-2896, doi:10.1109/TAP.2022.3233718 (2023).

25  Liu, K. *et al.* Orbital-Angular-Momentum-Based Electromagnetic Vortex Imaging. *IEEE Antennas and Wireless Propagation Letters* **14**, 711-714, doi:10.1109/LAWP.2014.2376970 (2015).

26  Ren, H. *et al.* Metasurface orbital angular momentum holography. *Nature Communications* **10**, 2986, doi:10.1038/s41467-019-11030-1 (2019).

27  Ni, J. *et al.* Multidimensional phase singularities in nanophotonics. *Science* **374**, eabj0039, doi:10.1126/science.abj0039 (2021).

28  Shen, Y. *et al.* Optical vortices 30 years on: OAM manipulation from topological charge to multiple singularities. *Light: Science & Applications* **8**, 90, doi:10.1038/s41377-019-0194-2 (2019).

29  Yang, H. *et al.* A THz-OAM Wireless Communication System Based on Transmissive Metasurface. *IEEE Transactions on Antennas and Propagation* **71**, 4194-4203, doi:10.1109/TAP.2023.3255539 (2023).

30  Chung, H., Kim, D., Choi, E. & Lee, J. E-Band Metasurface-Based Orbital Angular Momentum Multiplexing and Demultiplexing. *Laser & Photonics Reviews* **16**, 2100456, doi:https://doi.org/10.1002/lpor.202100456 (2022).

31  Barreiro, J. T., Wei, T.-C. & Kwiat, P. G. Beating the channel capacity limit for linear photonic superdense coding. *Nature Physics* **4**, 282-286, doi:10.1038/nphys919 (2008).

32  Wang, J. *et al.* Terabit free-space data transmission employing orbital angular momentum multiplexing. *Nature Photonics* **6**, 488-496, doi:10.1038/nphoton.2012.138 (2012).

33  Ren, H., Li, X., Zhang, Q. & Gu, M. On-chip noninterference angular momentum multiplexing of broadband light. *Science* **352**, 805-809, doi:10.1126/science.aaf1112 (2016).

34  Garcia-Gracia, H. & Gutiérrez-Vega, J. C. Diffraction of plane waves by finite-radius spiral phase plates of integer and fractional topological charge. *J. Opt. Soc. Am. A* **26**, 794-803, doi:10.1364/JOSAA.26.000794 (2009).

35  Chen, Y. *et al.* A Flat-Lensed Spiral Phase Plate Based on Phase-Shifting Surface for Generation of Millimeter-Wave OAM Beam. *IEEE Antennas and Wireless Propagation Letters* **15**, 1156-

1158, doi:10.1109/LAWP.2015.2497243 (2016).

36  Xie, Z. et al. Integrated (de)multiplexer for orbital angular momentum fiber communication. *Photon. Res.* **6**, 743-749, doi:10.1364/PRJ.6.000743 (2018).

37  Zhou, N. et al. Ultra-compact broadband polarization diversity orbital angular momentum generator with 3.6 × 3.6 μm2 footprint. *Science Advances* **5**, eaau9593, doi:10.1126/sciadv.aau9593 (2019).

38  Zhao, Z., Wang, J., Li, S. & Willner, A. E. Metamaterials-based broadband generation of orbital angular momentum carrying vector beams. *Opt. Lett.* **38**, 932-934, doi:10.1364/OL.38.000932 (2013).

39  Barati Sedeh, H., Salary, M. M. & Mosallaei, H. Time-varying optical vortices enabled by time-modulated metasurfaces. *Nanophotonics* **9**, 2957-2976, doi:doi:10.1515/nanoph-2020-0202 (2020).

40  Pujing Lin et al. Enabling Intelligent Metasurfaces for Semi-known Input. *Progress In Electromagnetics Research* **178**, 83-91 (2023).

41  Cao, X. et al. Digitized subwavelength surface structure on silicon platform for wavelength-/polarization-/charge-diverse optical vortex generation. *Nanophotonics* **11**, 4551-4564, doi:doi:10.1515/nanoph-2022-0395 (2022).

42  Tsuruda, K., Fujita, M. & Nagatsuma, T. Extremely low-loss terahertz waveguide based on silicon photonic-crystal slab. *Opt. Express* **23**, 31977-31990, doi:10.1364/OE.23.031977 (2015).

43  White, A. D. et al. Inverse Design of Optical Vortex Beam Emitters. *ACS Photonics* **10**, 803-807, doi:10.1021/acsphotonics.2c01007 (2023).

44  Dupačová, J., Consigli, G. & Wallace, S. W. Scenarios for Multistage Stochastic Programs. *Annals of Operations Research* **100**, 25-53, doi:10.1023/A:1019206915174 (2000).

45  Su, L. et al. Nanophotonic inverse design with SPINS: Software architecture and practical considerations. *Applied Physics Reviews* **7**, 011407, doi:10.1063/1.5131263 (2020).

46  Hamdan, B., Liu, Z., Ho, K., Büyüktahtakın, İ. E. & Wang, P. A dynamic multi-stage design framework for staged deployment optimization of highly stochastic systems. *Structural and Multidisciplinary Optimization* **66**, 162, doi:10.1007/s00158-023-03609-6 (2023).

47  Lalau-Keraly, C. M., Bhargava, S., Miller, O. D. & Yablonovitch, E. Adjoint shape optimization applied to electromagnetic design. *Opt. Express* **21**, 21693-21701, doi:10.1364/OE.21.021693 (2013).

48  Xiao, S. et al. Artificial bee colony algorithm based on adaptive neighborhood search and Gaussian perturbation. *Applied Soft Computing* **100**, 106955, doi:https://doi.org/10.1016/j.asoc.2020.106955 (2021).

49  Zhou, M., Lazarov, B. S., Wang, F. & Sigmund, O. Minimum length scale in topology optimization by geometric constraints. *Computer Methods in Applied Mechanics and Engineering* **293**, 266-282, doi:https://doi.org/10.1016/j.cma.2015.05.003 (2015).

50  Method of anisotropic etching of silicon. United States patent (2003).

51  He, X., Ren, B., Chan, K. F. & Wong, A. M. H. Full-Space Spin-Controlled Four-Channel Metalens With Equal Power Distribution and Broad Bandwidth. *Laser & Photonics Reviews* **19**, 2401843, doi:https://doi.org/10.1002/lpor.202401843 (2025).